\documentclass[fleqn,usenatbib]{mnras}

\usepackage{newtxtext,newtxmath}

\usepackage[T1]{fontenc}

\DeclareRobustCommand{\VAN}[3]{#2}
\let\VANthebibliography\thebibliography
\def\thebibliography{\DeclareRobustCommand{\VAN}[3]{##3}\VANthebibliography}

\usepackage{graphicx}	
\usepackage{amsmath}	
\usepackage{pdflscape}
\usepackage[dvipsnames]{xcolor}

\title[Driving RSG winds with binary interaction]{Driving asymmetric red supergiants winds with binary interactions}

\author[Landri \& Pejcha]{Camille Landri,$^{1}$\thanks{E-mail: \href{mailto:camille.landri@utf.mff.cuni.cz}{camille.landri@utf.mff.cuni.cz}}
Ond\v{r}ej Pejcha,$^{1}$ \\
$^{1}$Institute of Theoretical Physics, Faculty of Mathematics and Physics, Charles University, V Holešovičkách 2, 180 00 Praha 8, Czech Republic\\
}

\date{Accepted XXX. Received YYY; in original form ZZZ}

\pubyear{2024}

\begin{document}
\label{firstpage}
\pagerange{\pageref{firstpage}--\pageref{lastpage}}
\maketitle

\begin{abstract}
Massive stars in the red supergiant (RSG) phase are known to undergo strong mass loss through winds and observations indicate that a substantial part of this mass loss could be driven by localised and episodic outflows. Various mechanisms have been considered to explain this type of mass loss in RSGs, but these models often focus on single-star evolution. However, massive stars commonly evolve in binary systems, potentially interacting with their companions. 
Motivated by observations of the highly asymmetric circumstellar ejecta around the RSG VY~CMa, we investigate a scenario where a companion on an eccentric orbit grazes the surface of a red supergiant at periastron. The companion ejects part of the outer RSG envelope, which radiatively cools, reaching the proper conditions for dust condensation and eventually giving rise to dust-driven winds. Using simple treatments for radiative cooling and dust-driven winds, we perform 3D smoothed particle hydrodynamics simulations of this scenario with a $20\,M_\odot$ RSG and a $2\,M_\odot$ companion. We follow the evolution of the binary throughout a total of 14 orbits and observe that the orbit tightens after each interaction, in turn enhancing the mass loss of subsequent interactions. We show that one such grazing interaction yields outflows of $3\times10^{-4}\,M_\odot$, which later results in wide asymmetric dusty ejecta, carrying a total mass of $0.185\,M_\odot$ by the end of simulations. We discuss the implications for the evolution of the binary, potential observational signatures, as well as future improvements of the model required to provide sensible predictions for the evolution of massive binaries.
\end{abstract}




\section{Introduction} 
The red supergiant (RSG) phase is an important part of the evolution of massive stars with initial masses between $8$ and $25\,M_\odot$ \citep[e.g.,][]{ekstrom_grids_2012}, during which they undergo substantial mass loss through winds. Observations show that a significant part of this mass loss can be driven by episodic outflows localised on the stellar surface \citep{humphreys_mass-loss_2021}, as is shown by the ejecta surrounding VY~CMa. VY~CMa is a RSG located at 1.2\,kpc \citep{zhang_distance_2012} and is one of the largest and most massive RSGs observed to this date, with a current radius of $1420\,\pm\,120R_\odot$, a luminosity of $(2.7\pm0.4)\times 10^5 L_\odot$, and an initial mass at the upper range for RSGs \citep[$25-32\,M_\odot$,][]{wittkowski_fundamental_2012}. Various observations of VY CMa show that it is embedded in asymmetric circumstellar ejecta with distinct and complex structures such as arcs, knots, and clumps \citep{smith_asymmetric_2001, humphreys_three-dimensional_2007, jones_three-dimensional_2007, kaminski_interferometric_2013, richards_alma_2014, ogorman_alma_2015, decin_alma-resolved_2016, vlemmings_magnetically_2017, gordon_thermal_2019,kaminski_massive_2019,humphreys_mass-loss_2021,quintana-lacaci_history_2023,humphreys_hidden_2024}. The clumps in particular are thought to be caused by highly localised mass-loss events that have been occurring for the past 1200~yrs \citep{shenoy_searching_2016}, but their origin so far remains unexplained.

Various mechanisms have been considered to explain episodic mass loss in single RSGs, generally involving dust-driven winds launched by a disturbance of the stellar surface by magnetic activity, convection, or pulsations (e.g., \citealt{smith_asymmetric_2001,humphreys_three-dimensional_2007,ogorman_alma_2015,vlemmings_magnetically_2017}). However, quantitative prescriptions of these processes are scarce and often ambiguous, and stellar evolution models of cool supergiants continue to use time-averaged empirical mass loss formulations that do not take into account the episodic nature of RSG mass loss \citep[e.g.,][]{de_jager_mass_1988,nieuwenhuijzen_parametrization_1990,van_loon_empirical_2005}. Since these localised outflows can represent a large fraction of the mass lost by the RSG, as is the case for VY~CMa, our poor understanding of these mechanisms introduces large uncertainties in current massive star evolution models \citep[e.g.,][]{smith14} and serious discrepancies between theory and observations (e.g., \citealt{massey_time-averaged_2023}). Furthermore, a lot of important astrophysical processes, such as gravitational waves emission from compact binaries, chemical evolution of galaxies, and core-collapse supernovae (SNe), strongly depend on massive star evolution. It is therefore crucial to improve our understanding of episodic mass loss in RSGs and better constrain their fate.

An important aspect of RSG evolution is multiplicity. A significant fraction of massive stars are found in binary systems \citep[e.g.,][]{mason_high_2009, sana_binary_2012,moe_mind_2017}, with a large variety of possible configurations, including some where the separation of the two stars is small enough to allow them to interact. A small but significant part of these systems are expected to remain bound and interact throughout their whole evolution, leading to the formation of objects such as X-ray binaries or gravitational wave events. For instance, common formation channels for double neutron stars binaries show multiple evolution stages where one of the star is a RSG interacting with its companion \citep[e.g.,][]{tauris_formation_2017}. Therefore, scenarios where binary interaction and stellar winds interplay are likely to occur. While the impact of binarity on stellar winds has been investigated in the case of stars on the Asymptotic Giant Branch (AGB) \citep[e.g.,][]{chen_3d_2020,bermudez-bustamante_agb_2020,aydi_3d_2022, esseldeurs_3d_2023}, it remains widely unexplored for RSG stars. 

\begin{figure}
    \centering
    \includegraphics[width=\columnwidth]{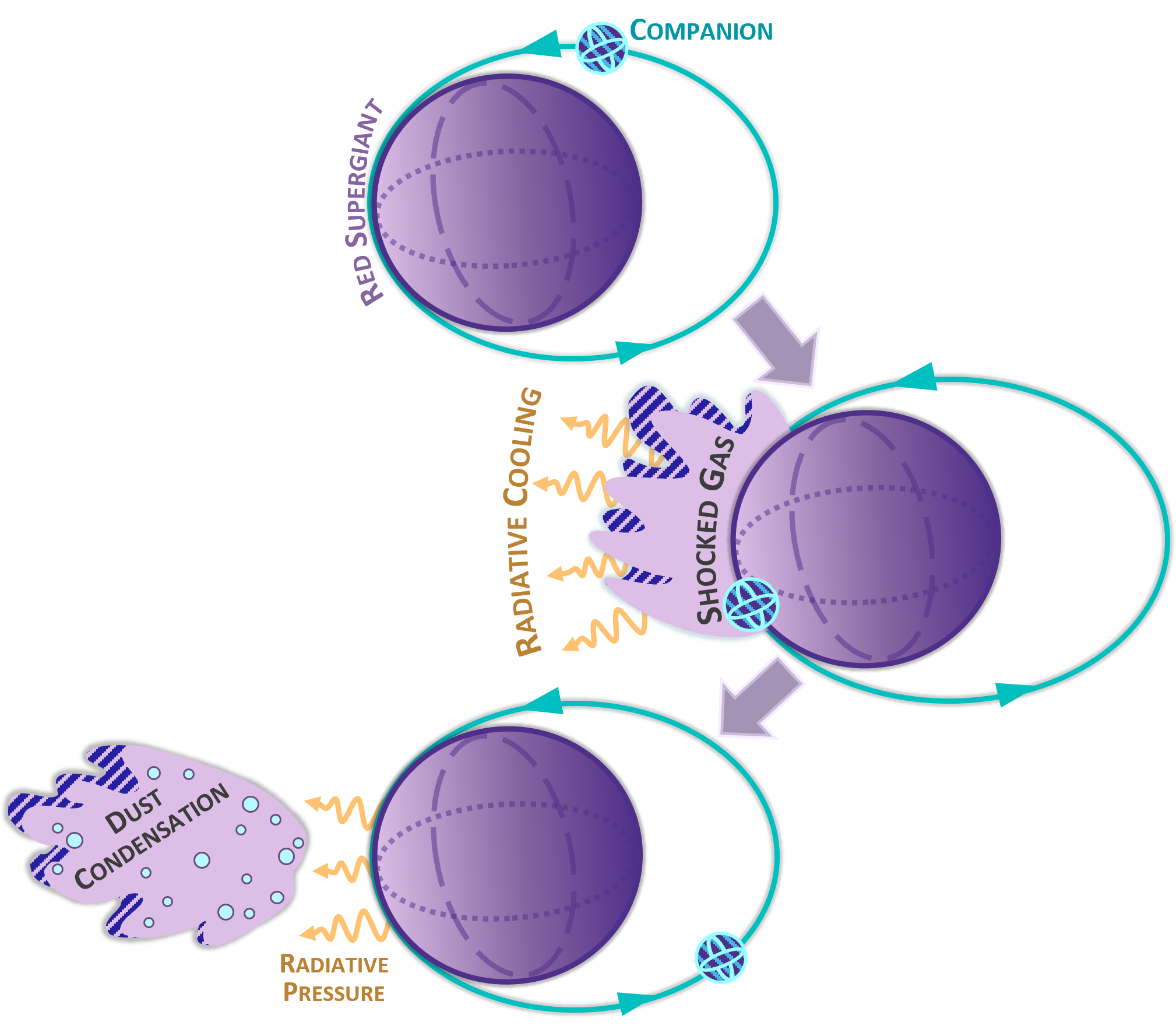}
    \caption{Schematics of the scenario explored with our simulations. A companion on an eccentric orbit grazes the envelope of the red supergiant and the resulting outflow radiatively cools, eventually reaching temperatures that permit dust formation. Radiation pressure then accelerates the dust grains, which drag the gas along, resulting in asymmetric dust-driven winds.}
    \label{fig:setup}
\end{figure}

In this paper, we investigate a scenario where binary interaction can drive anisotropic mass loss episodes in RSGs that can later turn in dust-driven winds. More specifically, we consider a RSG with a companion grazing the RSG envelope on a highly eccentric orbit. We illustrate this scenario in Fig.~\ref{fig:setup}: the companion grazes the outer envelope at each periastron passage and the shocked gas is ejected from the envelope. As the ejected gas spreads outward, it cools and eventually reaches temperatures that are low enough to allow dust to condense. Radiative pressure then accelerates the newly formed dust grains, dragging the gas along and effectively driving an asymmetric wind. Eventually, the companion enters deep into the envelope of the RSG and commences a common envelope evolution \citep[e.g.,][]{paczynski_common_1976,fragos19, ropke_simulations_2023,lau_common_2022,gagnier23,gagnier24}. 
This scenario resembles the grazing envelope evolution proposed by \cite{soker_close_2015} except we do not involve accretion disk and jets around the companion and the resulting outflow morphology is different \citep{shiber17}. Our model makes use of radiation pressures on dust grains, which was suggested as important in late stages of common envelope evolution by \citet{glanz18}. \citet{glanz21} also studied common envelope evolution in eccentric binaries, but did not address anisotropic outflows during periastron passages.

We explore the proposed scenario through 3D hydrodynamics simulations, where we include a simplified treatment of dust formation and radiation pressure driving the RSG outflows. We note that the aim of this proof-of-concept study is not to reproduce the detailed properties and complicated morphology of the ejecta surrounding VY~CMa or other RSGs, where in reality they likely arise from an interplay of many different processes. Instead, our goal is to illustrate that binary interactions can potentially explain some of the observed features. A more sophisticated treatment of dust formation and radiative processes can be added in follow-up studies to reach a better agreement with the observations.
In Section~\ref{sec:methods}, we describe the methods used to model our scenario, including how we simplified the different treatments of the physical processes involved. In Section~\ref{sec:results}, we present our results, including evolution of the orbit, amounts of mass ejected, and  the ejecta expansion. In Section~\ref{sec:discussion}, we discuss the possible formation channels for our system ,the implications of our model for binary evolution evolution, the possible observational signatures of such winds, and future improvements for our models.

\section{Methods}\label{sec:methods}

We use the Smoothed Particle Hydrodynamics (SPH) code \textsc{Phantom} \citep{price_phantom_2018} that solves the Lagrangian form of the equations of hydrodynamics by discretising the fluid as particles with mass $m$ and local fluid velocity $v$ \citep[e.g.,][]{lucy_numerical_1977, gingold_smoothed_1977,price_smoothed_2012}.
Physical quantities are then computed by summing particle contributions weighted by a smoothing kernel $W$, for instance the density around particle $a$ is given by:
\begin{equation}\label{eq:densitysum}
    \rho_a=\sum_bm_bW\left(\lvert\vec{r}_a-\vec{r}_b\rvert,h_a\right),
\end{equation}
where $h_a$ is the smoothing length, which defines the neighbourhood of the particle $a$ and is proportional to the local particle number density $h_a=h_\textrm{fact}n_a^{-1/3}$, $h_\textrm{fact}$ is a proportionality factor. The resolution of the simulations is therefore set by the total number of particles used, the choice of smoothing kernel and $h_\textrm{fact}$.

For this work, we set up \textsc{Phantom} is set up to solve the equations of hydrodynamics in the following form:
\begin{align}
    \frac{d\vec{v}}{dt} =& -\frac{\nabla P}{\rho}+\Pi_\textrm{shock}+\vec{a}_\textrm{selfgrav}+\vec{a}_\textrm{ext},
\label{eq:force} \\
    \frac{du}{dt}=&-\frac{P}{\rho}\left(\nabla\cdot\vec{v}\right)+\Psi_\textrm{shock}-\frac{\Upsilon_\textrm{cool}}{\rho},
\label{eq:ener}
\end{align}
where $P$ is the pressure, $u$ is the internal energy, $\vec{a}_\textrm{selfgrav}$ is the acceleration due to self-gravity, $\vec{a}_\textrm{ext}$ represents the acceleration due to external forces such as sink particles or a radiative flux, $\Pi_\textrm{shock}$ and $\Psi_\textrm{shock}$ are viscous dissipation terms, and $\Upsilon_\textrm{cool}$ is an optional cooling term. For $W$, we use the M$_4$ cubic spline kernel with $h_\text{fact}=1.2$, corresponding to an average of 58 neighbours . 

Our scenario involves more than hydrodynamics and self-gravity. As shown in Fig~\ref{fig:setup}, our simulations require a treatment of radiative cooling as the shocked gas is ejected from the RSG, and a treatment of dust condensation and radiative pressure to accelerate the outflow. Since this work is meant to be a proof-of-principle, these treatments will remain simple, but we do plan on improving them in follow-up studies. For this study, we run a set of 4 simulations which are summarised in Tab.~\ref{tab:runs}. In the rest of this section, we describe how we initialise the RSG and the binary system (Sec.~\ref{sec:init}), as well as how we treat dust-driven winds (Sec.~\ref{sec:dust}) and radiative cooling (Sec.~\ref{sec:cooling}).

\begin{table} 
\centering
\caption{Summary of the simulations performed for this study, showing the number of particles, minimum softening length $h_\text{a}$, and processes involved in the run.\label{tab:runs}} 
\begin{tabular}{c|cp{3.5cm}|}
\hline \hline
Number of particles & $\min h_\text{a}$ & Physics involved \\
\hline 
$2.5\times10^5$     & 22~$R_\odot$ & Hydro, self-gravity, cooling, dust-driven winds \\ 
$1\times10^6$       & 14~$R_\odot$ & Hydro, self-gravity, cooling, dust-driven winds\\
$1\times10^6$       & 14~$R_\odot$ & Hydro, self-gravity,  adiabatic, no wind\\
$2\times10^6$       & 11~$R_\odot$ & Hydro, self-gravity, cooling, dust-driven winds\\
\hline\hline
\end{tabular}
\end{table}

\subsection{Initialisation of the system}\label{sec:init}
We consider a binary of mass ratio $q=M_2/M_1=0.1$ with a $M_1=20\,M_\odot$ RSG similar to VY~CMa, and a companion of $M_2=2\,M_\odot$, which could represent either a low-mass non-degenerate star or a fairly massive neutron star (NS) \citep[e.g.,][]{ozel_masses_2016}. We initialise this system in two steps, first the 1D stellar profile of the RSG is mapped to 3D and relaxed in \textsc{Phantom}. During this phase, there are no external accelerations and no cooling. Then, after ensuring the stability of the 3D model, we add the companion and let the binary evolve for several orbital periods. 

For interior of the RSG, we create a 1D stellar profile with properties similar to VY~CMa, which has mass and radius estimated to be $17\pm8\,M_\odot$ and $1420\pm120\,R_\odot$ respectively \citep{wittkowski_fundamental_2012}. Since the companion mostly interacts with the outer layers of the RSG, it is unnecessary and computationally expensive to resolve the core and inner envelope of the giant. We therefore replace part of the stellar interior with a sink particle, a point mass that interacts with other particles only gravitationally through a potential smoothed with a cubic spline kernel. We excised about half of the envelope, which increases the number of particles at the stellar surface for a reasonable total number of particles as well as maintains a reasonable timestep during the simulations. For the rest of the envelope we choose to use an artificial RSG profile that is convectively stable, since reproducing accurate convective motion requires to resolve the inner envelope and we wish to avoid the decrease in timestep associated with these dense envelope layers. To create the model for the envelope, we follow the method provided in Appendix~A of \cite{lau_common_2022} and solve the equations of hydrostatic equilibrium with an ideal gas equation of state assuming an adiabatic index of $\gamma=5/3$, constant entropy, and accounting for the softened potential of the core. To solve these equations, we used boundary conditions from a realistic RSG profile obtained with \textsc{Mesa} \citep{paxton_modules_2011,paxton13,paxton15,paxton18,paxton19,jermyn23} made to match the properties of VY~CMa (for a detailed description of the process see Appendix~\ref{ap:rsgprof}).

\begin{figure}
    \centering
    \includegraphics[width=\columnwidth]{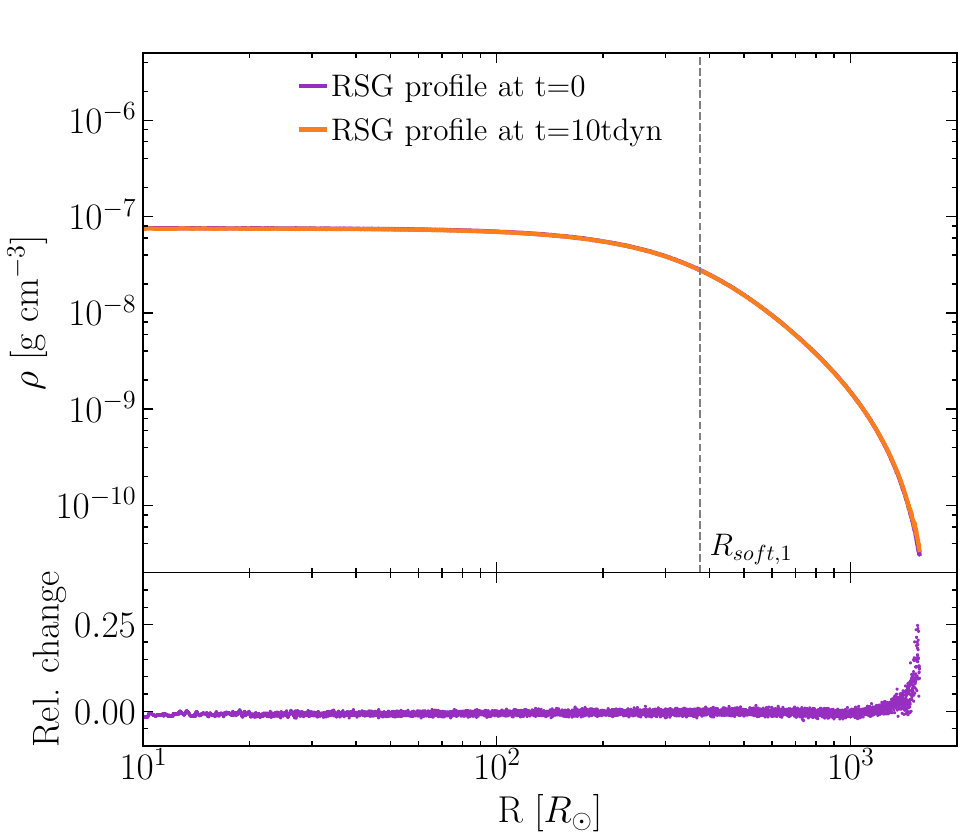}
    \caption{Density profile of the outer envelope of the RSG in our simulation at the end of the relaxation and 10~$\tau_\text{dyn}$ later. Top: Density profiles, purple shows the initial profile and orange the profile after 10~$\tau_\text{dyn}$, the grey dashed line shows the softening radius of the RSG numerical core. Bottom: Relative change in density computed as $(\rho_\text{f}-\rho_\text{i})/\rho_\text{i}$. The envelope remains stable apart from a 25\% increase in density at the surface.}
    \label{fig:dens}
\end{figure}

Following this process, we created a 1D convectively stable profile of a $20\,M_\odot$ and $1500\,R_\odot$ RSG that is shown in Fig.~\ref{fig:dens}. We initialised it in \textsc{Phantom} with a sink particle core of 13.75\,$M_\odot$ and a potential smoothed with a cubic spline kernel. We define the smoothing length of this potential to be $R_\text{soft,1}=375\,R_\odot$, yielding an effective radius of 750\,$R_\odot$ for the sink particle since the Newtonian potential is recovered at $2r_\text{soft}$ with a cubic spline kernel. The 1D stellar profile of the RSG is mapped to a 3D distribution and then relaxed using the procedure implemented in \textsc{Phantom} by \cite{lau_common_2022}. The giant is then evolved for 10 dynamical timescales ($\tau_\text{dyn}=241\,d$) to ensure the stability of the star. In Fig.~\ref{fig:dens}, we show the density profile before and after relaxation: while the RSG remains stable, the relaxation and subsequent evolution yield a small change in stellar radius as the RSG has slightly expanded during the relaxation and then contracted. For the highest resolution simulation ($2\times 10^6$ particles) the RSG has expanded from $1500\,R_\odot$ to $1569\,R_\odot$.

Once the star is relaxed and stable, we add the neutron star companion by placing it at the apoastron of the orbit. Since a neutron star is much smaller than the minimum smoothing length in our simulations, we initialise it as a sink particle of mass $M_2=2\,M_\odot$ with smoothing length and accretion radius equal to the minimum resolved length, which is  $R_\text{soft,2}=11\,R_\odot$ for our highest resolution simulations with $2\times 10^6$ particles. The orbit of the system is set with a semi-major axis $A=7500R_\odot$ and eccentricity $e=0.8$, so that the secondary grazes the envelope of the RSG at each periastron passage. We then evolve the system for several orbits until the companion has plunged deeply into the RSG envelope and the smoothed potential of the two sink particles overlap.

\begin{table} 
\centering
\caption{Initial conditions of the high resolution simulations ($2\times 10^6$ particles). $T_\text{cond}$ and $p_\text{dust}$ are the dust condensation parameters according to equations~(\ref{eq:temp}) and (\ref{eq:opac}). \label{tab:setup}} 
\begin{tabular}{l|ll|}
\hline \hline
RSG primary & $M_1$ & 20.0 $M_\odot$ \\ 
            & $R_1$ & 1569 $R_\odot$ \\
            & $R_\text{soft,1}$ & 375 $R_\odot$\\
            & $M_\text{core}$ & 13.75 $M_\odot$ \\
            & $M_\text{env}$ & 6.25 $M_\odot$ \\
            & $T_\text{eff}$ & 3500 K \\\hline
Companion   & $M_2$ & 2.0 $M_\odot$  \\
            & $R_\text{soft,2}$ & 11 $R_\odot$\\ \hline
Binary      & Eccentricity & 0.8 \\  
            & Semi-major axis & 7500 $R_\odot$ \\
            & Period & 43.91 yrs \\   \hline
Dust        & $T_\text{cond}$ & 1100 K\\
            & $p_\text{dust}$ & -0.9 \\
\hline\hline
\end{tabular}
\end{table}

\subsection{Cooling}\label{sec:cooling}
As the companion goes through the envelope of the giant, the shocked gas that is ejected becomes optically thin and cools radiatively to the local equilibrium temperature $T_\text{eq}$, where radiative cooling and irradiation from the RSG compensate each other. Under the assumptions of spherical symmetry and that the RSG radiative intensity dominated the local intensity, $T_\text{eq}$ is given by \citep{lamers_introduction_1999}:
\begin{equation}\label{eq:temp}
    T_\text{eq}= T_{\text{eff},1} W(\tilde{r})^{1/(4+p_\text{dust})}, \\
    W(\tilde{r}) = \frac{1}{2}\left ( 1- \sqrt{1 - \left (\frac{R_1}{\tilde{r}} \right )^2}  \right )
\end{equation}
where $T_{\text{eff},1}$ and $R_1$ are the  effective temperature and radius of the RSG, $\tilde{r}$ is the distance from the center of mass of the RSG, and $W(\tilde{r})$ is the so-called geometrical dilution factor. Since we are primarily interested in situations where $T_\text{eq}$ decreases below the dust condensation temperature$T_\text{cond}$, equation~(\ref{eq:temp}) includes dust correction through the exponent $p_\text{dust}$, which comes from approximating the wavelength dependent part of the dust opacity with a power law,
\begin{equation}\label{eq:opac}
    \kappa_\text{d}= \kappa_0  \left (\frac{\lambda}{\lambda_0} \right )^{-p_\text{dust}}.
\end{equation}
The dust properties are described in Section~\ref{sec:dust}.


Since cooling impacts the kinematics of the ejected gas, it is important to take it into account in our simulations. The most natural way to handle this process would of course be to fully resolve radiative cooling and irradiation by imposing a cooling timescale over which the gas relaxes to $T_\text{eq}$. 
The cooling timescale $t_\text{cool}$ over which a shocked gas radiates away its internal energy $U=3/2k_\text{B}T$ can be estimated from a radiative cooling function (e.g., Fig.~22 of \citealt{ferland_2017_2017}),
\begin{equation}\label{eq:cool}
    t_\text{cool}=\frac{U m_\text{p}}{\rho \Lambda(T)}.
\end{equation}
Here, $m_\text{p}$ is the proton mass, $\rho$ is the density of the gas, and $\Lambda(T)$ is the usual cooling function. As the companion goes through the envelope of the giant, the shocked upper layers of the envelope reach temperatures up to 35000~K, for which the cooling rate is approximately $3.16\times10^{-4}$~erg~cm$^3$~s$^{-1}$. For a density of $10^{-10}$~g~cm$^{-3}$, equation~(\ref{eq:cool}) yields $t_\text{cool} \approx 3.2\times10^{-4}$~s. Considering the hydrodynamical timestep $t_\text{step}$ of our simulations is of the order of hours, it is impossible to resolve such fast cooling in our simulations without making their cost unreasonably high.

We therefore have to resort to a simpler method to cool the ejected gas. Instead of imposing a timescale for the cooling process, the ejected particles are cooled so that they would exponentially reach $T_\text{eq}$ on a timescale equal to the hydrodynamical timestep. This ensures that cooling does not become faster than the explicit timestep and should prevent development of cooling instabilities. To do so, the equilibrium temperature at a particle location $T_\text{eq,a}$ is calculated using equation~(\ref{eq:temp}), and the corresponding specific internal energy is obtained using the ideal gas equation of state:
\begin{equation}
    u_{\text{eq},a} = \frac{k_\text{b} T_{\text{eq},a}}{(\gamma - 1) \mu m_\text{p} },
\end{equation}
where the adiabatic index is $\gamma=5/3$ and the mean molecular weight is $\mu=0.659$. The local cooling rate per unit volume is then calculated as:
\begin{equation}
    \Upsilon_{\text{cool},a} = \rho \frac{u_{a}-u_{\text{eq},a}}{t_\text{step}},
\end{equation}
where $u_\text{a}$ is the specific internal energy of the particle before cooling is applied. This cooling rate is applied in equation~(\ref{eq:ener}) for any particle that is considered ``ejected'', i.e. is outside of the RSG envelope,
\begin{equation}
    \Upsilon_\text{cool} = 
    \begin{cases}
    \Upsilon_{\text{cool},a} & \text{if} \quad \tilde{r}_{a}>R_1, \\
    0 & \text{if} \quad \tilde{r}_{a}\leq R_1.
    \end{cases} 
\end{equation}
To avoid approaching too low temperatures, we apply an arbitrary floor temperature of 500\,K. Overall, our treatment of the radiative cooling process is an oversimplification and will require more accurate treatment in follow-up studies.

\subsection{Dust-driven winds}\label{sec:dust}
In our scenario, the wind is driven by the radiation pressure on the dust condensing in the ejecta lifted by the companion passage. As the dust grains are accelerated outward by the radiative flux, they drag the gas along by transferring momentum through collisions. Accurately reproducing the formation of such winds therefore requires to resolve dust formation (and destruction), the radiative acceleration of dust grains, and dust-gas interactions. As this work aims to be a proof-of-concept, we will treat this problem with simple methods that should yield qualitatively similar outcome to more complete physical treatments. We leave the improvement of the wind treatment for follow-up studies.

The radiative acceleration of the dust grains due to the radiative flux of the RSG is taken into account by setting $\vec{a}_\text{ext}$ in equation~(\ref{eq:force}) so that it depends on a local Eddington factor $\Gamma_{a}$,
\begin{equation}
    \vec{a}_\text{ext,rad}= \frac{GM_1}{r_1^2}\Gamma_\text{a} \hat{r}_a,
\end{equation}
where $\hat{r}_a$ is the unit vector connecting particle $a$ to the RSG. The simplest treatment of stellar winds is the so-called free wind approximation \citep{theuns_wind_1993}, which consists of setting $\Gamma_{a}=1$ so that all particles escape the gravitational pull of the star. We apply this method to our simulations, adding the condition that only particles with properties fulfilling dust condensation criteria are accelerated. To find out whether dust can condense in a specific region, we compare the condensation temperature of the dust $T_\text{cond}$ to the temperature of the particle $T_{a}$ and set the value of $\Gamma_a$ as
\begin{equation}\label{eq:gamma}
    \Gamma_\text{a} = 
    \begin{cases}
    1 & \text{if} \quad T_{a} < T_\text{cond}, \\
    0 & \text{if} \quad T_{a} > T_\text{cond}.
    \end{cases} 
\end{equation}
We set the dust properties following \citet{bladh_exploring_2012} who parameterised dust-driven winds for different types of dust grains to determine the main wind-drivers. Based on their results and the low C/O ratio of RSG stars, we consider silicate grains composed of Mg$_2$SiO$_4$, for which they derived $T_\text{cond}=1100\,K$ and $p_\text{dust}=-0.9$. The dust parameters  We note that with this formalism, we do not account for interaction between the dust and gas, which are effectively fully coupled.

Overall, our method is similar to the procedure devised by \cite{bowen_dynamical_1988}, which has been widely used to treat dust condensation in AGB stellar winds \citep[e.g.,][]{bermudez-bustamante_agb_2020,chen_3d_2020,aydi_3d_2022, esseldeurs_3d_2023} and common envelope evolution \citep[e.g.,][]{gonzalez.2022,bermudez.2024}, but it differs in some aspects. Their simulations are always adiabatic, so they compare the local equilibrium temperature (calculated using equation~[\ref{eq:temp}]) to the condensation temperature, whereas in our simulations the gas cools as it is ejected, and we can directly use the gas temperature for the comparison. Additionally, they use this comparison to calculate a local Eddington factor $\Gamma$ that depends on the temperature difference and assumptions on gas and dust opacities, while we simply apply $\Gamma=1$ wherever dust condensation conditions are met.


\section{Results} \label{sec:results}
 
\begin{figure*}
    \centering
    \includegraphics[width=1\linewidth]{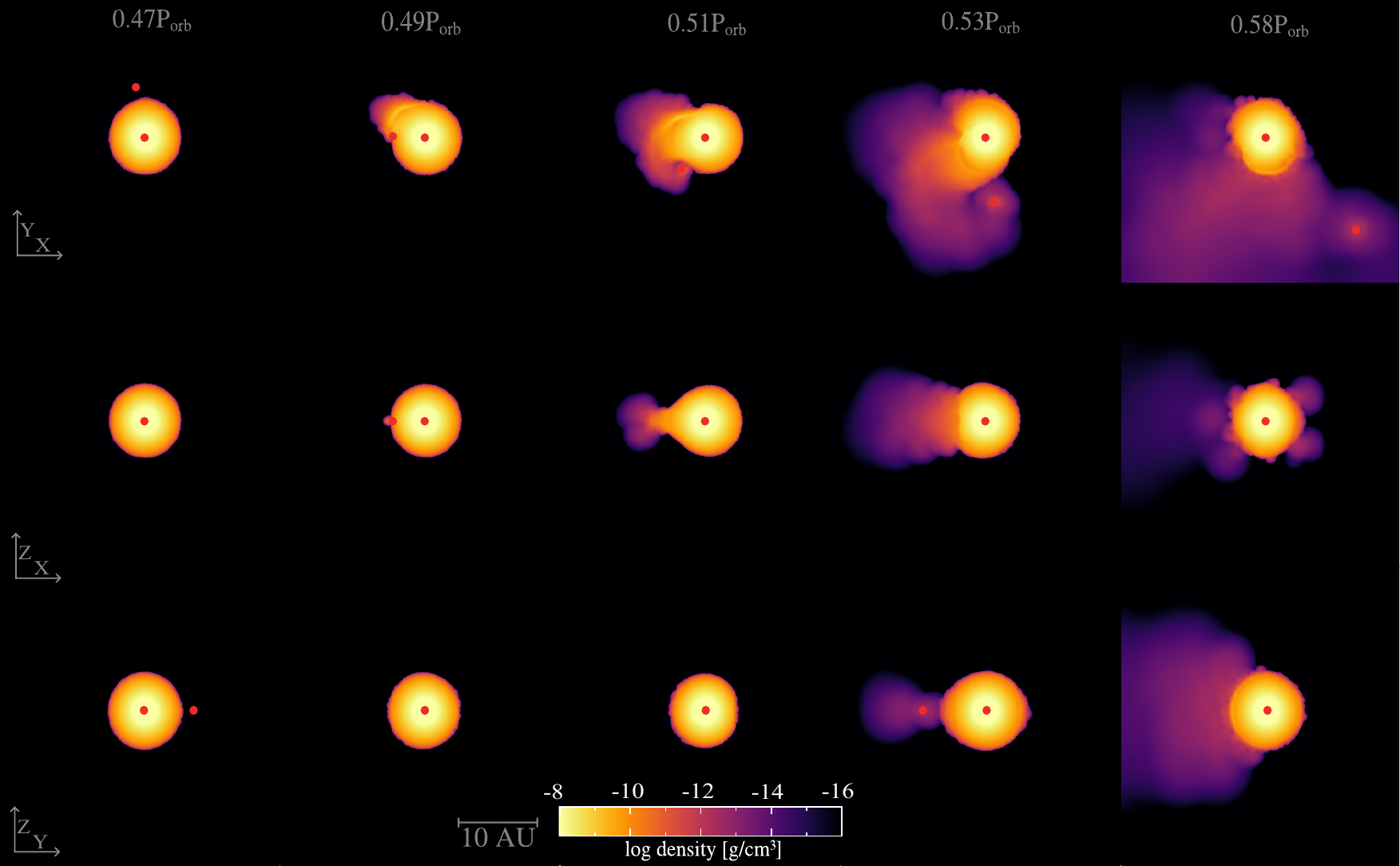}
    \caption{Snapshots of the density during the first periastron passage in our highest resolution run ($2\times 10^6$ particles). Upper, middle, and lower panels show cross-sections of density taken at $z=0$, $y=0$, and $x=0$ planes, respectively. Sink particles are denoted as red circles. A full movie of the first interaction can be found at \href{https://youtu.be/jcW0KyMayBE}{https://youtu.be/jcW0KyMayBE}}.
    \label{fig:snapclose}
\end{figure*}

We performed a total of 4 simulations: 3 runs with cooling, dust driving and varying resolution ($2.5\times10^5$, $1\times10^6$, and $2\times10^6$ particles), and one control adiabatic run without cooling and dust-driven winds with $1\times10^6$ particles. The parameters of the runs are summarised in Tables~\ref{tab:runs} and \ref{tab:setup}. 

\begin{figure}
    \centering
    \includegraphics[width=\columnwidth]{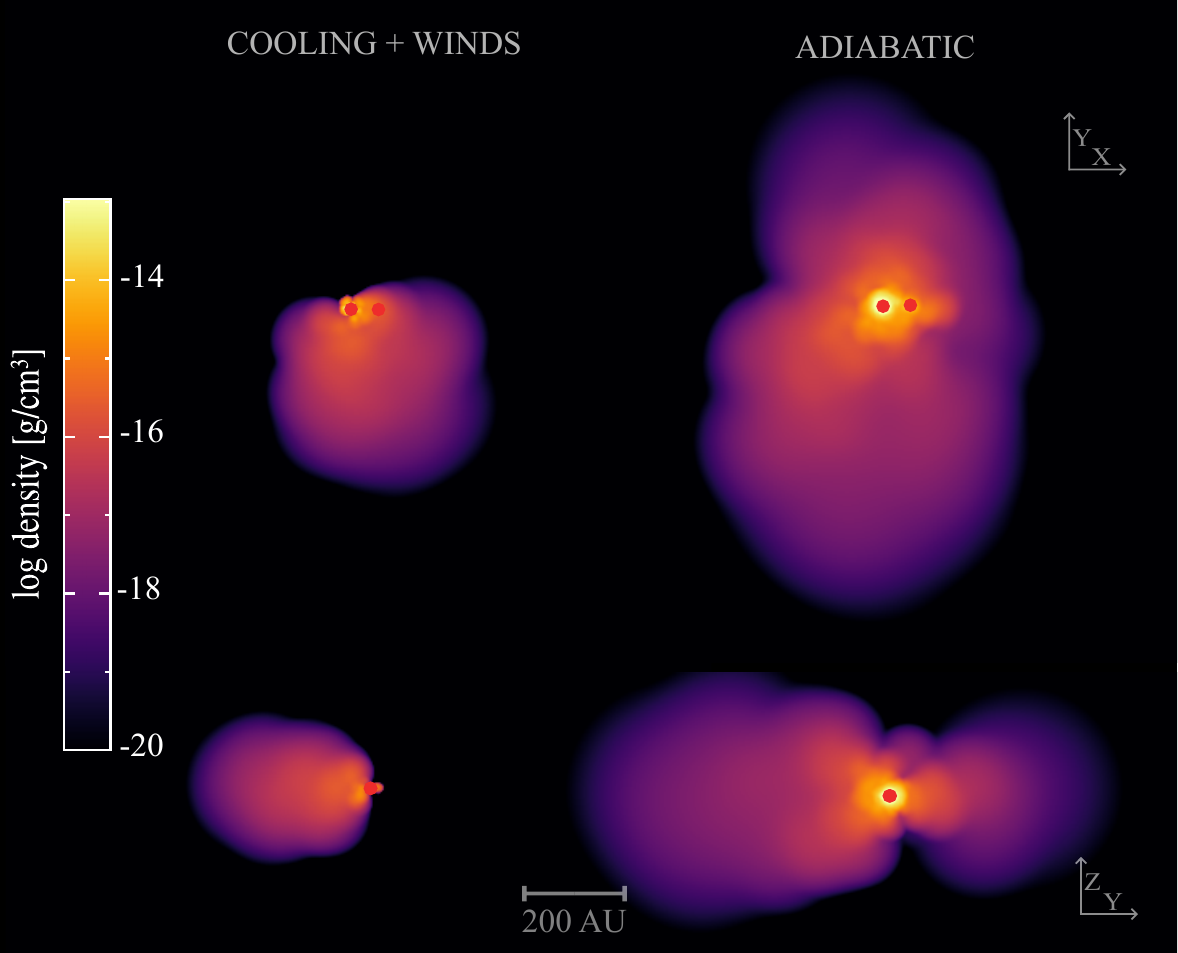}
    \caption{Comparison of the ejecta after one apastron passage of the companion for the adiabatic run (right panels) and a run with radiative cooling and dust-driven winds (left panels). Upper and lower panels show a cross section of the density taken along the $z=0$ and $x=0$ planes, respectively. Both simulations were performed with $1\times10^6$ particles.
    }
    \label{fig:comp}
\end{figure}

In Fig.~\ref{fig:snapclose}, we show snapshots of the density cross-section in our highest resolution simulation during the first grazing of the RSG envelope. As the companion grazes the outermost layers of the envelope, the shocked gas is ejected approximately perpendicularly to the stellar surface. However, it is then dragged by the companion as it leaves the vicinity of the RSG. Eventually, the outflow expands approximately radially in the $y<0$ region while we observe no ejection of gas in the $y>0$ region. The outflow expands and cools, reaching the dust condensation temperature $T_\text{cond}=1100$\,K for $r\gtrsim4500\,R_\odot$, where the gas density is about $10^{-14}$~g~cm$^{-3}$. Radiative pressure then starts to accelerate the dust grains, effectively supporting the radial expansion of the gas. During the following orbit, the outflow continues expanding while the RSG, which first expanded as a response to the perturbation from companion, is restoring its hydrostatic equilibrium. After one full orbit (around 40 years), the outflow has extended to a  rough semicircle of radius $r\sim400$\,au in the $xy$ plane with a thickness of $z\sim200$\,au, reaching densities as low as $10^{-18}$~g~cm$^{-3}$ and terminal velocities of around $40$~km~s$^{-1}$. Besides producing these asymmetric winds, we also expect the interaction between the companion and the RSG envelope to impact the orbit of the system, which in turn impacts the conditions of each subsequent grazing.

To assess the relevance of the cooling and the dust-driven winds in supporting the asymmetric outflows, we performed a fully adiabatic simulation. In Fig.~\ref{fig:comp}, we show the outflow resulting from one grazing interaction after the companion has passed the apastron in both the adiabatic and non-adiabatic simulations. We see that the ejecta in the adiabatic simulation is much more extended: it reaches $r\gtrsim500$\,au with a thickness of $\sim$300\,au, while the radiatively cooled ejecta has only spread to $r\sim200$~au with a thickness of $\sim$200\,au. Additionally, the adiabatic outflow is denser, reaching up to $10^{-14}$~g~cm$^{-3}$ in the inner region, against $10^{-15}-10^{-16}$~g~cm$^{-3}$ for the cooled outflow. Finally, the asymmetry of the ejecta with respect to the binary is much less pronounced in the adiabatic case: the adiabatic outflow broadly surrounds the RSG, while the gas is only ejected in a roughly semi-circular slab in the simulations with cooling and winds. This comparison shows that the cooling and wind prescriptions are essential to produce strongly asymmetric outflows. 

\subsection{Evolution of the orbit} 
\label{sec:orbit_evolution}
\begin{figure}
    \centering
    \includegraphics[width=\columnwidth]{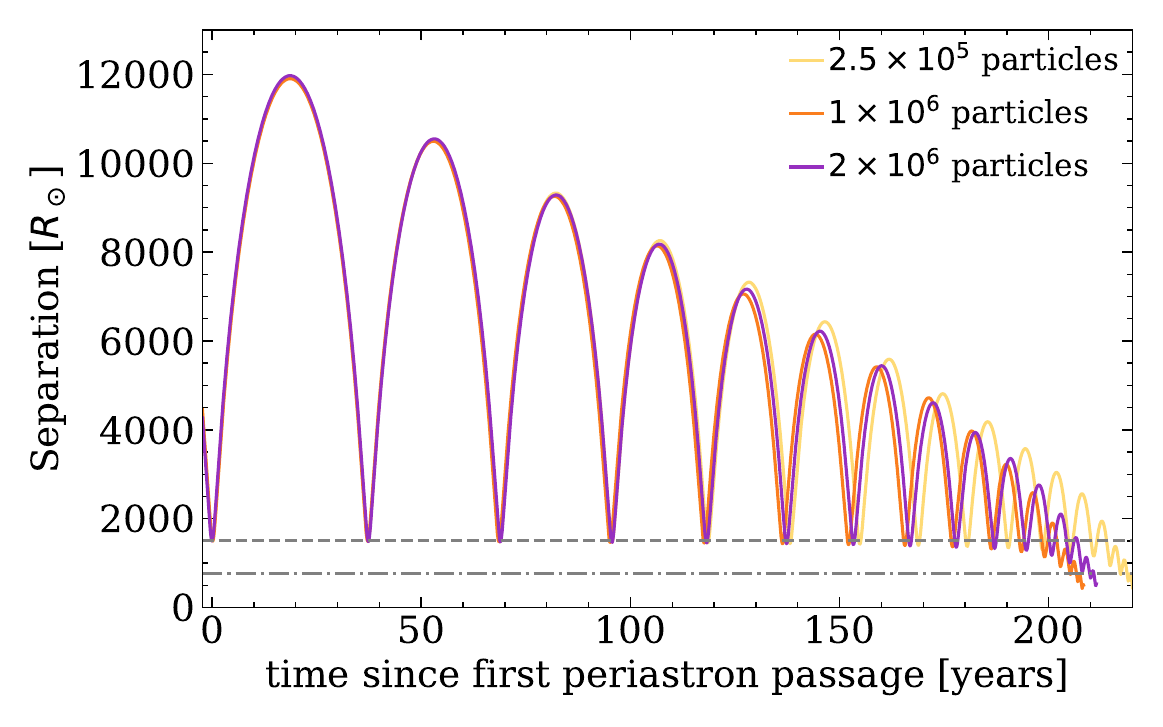}
    \caption{Evolution of the binary separation. The high, medium, and low resolution runs are shown in purple, orange, and yellow, respectively. The upper and lower dashed lines show the RSG surface ($1500\,R_\odot$) and the sum of the effective radii of the two sink particles for the high resolution run ($772\,R_\odot$).}
    \label{fig:sep}
\end{figure}

As the companion grazes the RSG envelope, we expect the drag between the gas and the companion to considerably affect the orbit of the binary. In Fig.~\ref{fig:sep}, we show the evolution of the separation of the binary. We see that during each subsequent periastron passage the companion reaches deeper layers of the outer envelope and the orbit tightens while remaining eccentric. The binary starts with a $\sim$40~yr period, which decreases on average by $3-4$~years after each orbit, reaching a period of about 7~years on the 10$^\text{th}$ orbit. At this point, the companion is more than $250\,R_\odot$ beneath the RSG surface at periastron, driving more massive and less localised outflows, and decreasing the orbital period even faster. 

After 13 orbits, the companion is fully engulfed by the RSG and starts spiraling in the envelope with period of the order of one year, effectively starting a phase of common envelope evolution (CEE) \citep[e.g.,][]{paczynski_common_1976,ropke_simulations_2023}. Unfortunately, we cannot resolve the inspiral of the companion in our simulations: the potential of the RSG sink core is smoothed up to $750\,R_\odot$, so we barely observe one full orbit of the inspiral before the smoothed potentials of the sink particles overlap, at the end of the 13$^\text{th}$ orbit. We expect that more mass will be ejected during CEE, and the outflow should retain some polar asymmetry until the CEE circularises the orbit. It is however not possible to determine the outcome of the CEE with our simulations. 

Overall, the orbit of the binary tightens drastically over the course of the simulation due to the grazing interaction, leading to CEE after only 200 years. We discuss implications and possible modifications to this timescale in Sec.~\ref{sec:discussion}.

\subsection{Mass loss}
\label{sec:massloss}

\begin{figure}
    \centering
    \includegraphics[width=\columnwidth]{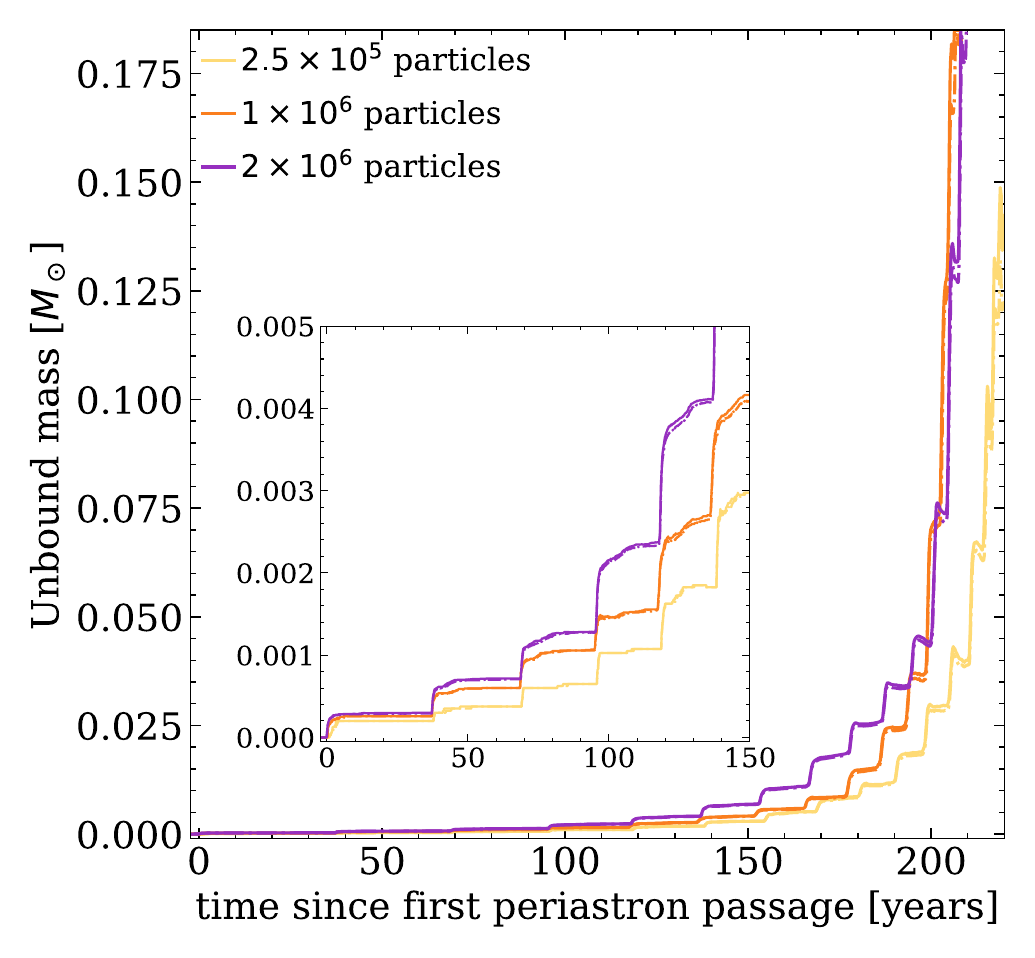}
    \caption{Evolution of the amount of unbound mass. The high, medium and low resolution runs are shown in purple, orange and yellow respectively. Plain lines show the amount of unbound mass according the energy criterion in equation~(\ref{eq:boundth}), which includes the gas internal energy, dashed lines show the amount of unbound mass using the more conservative energy criterion in equation~(\ref{eq:bound}), which does not include internal energy. The inner plot shows the mass loss during the first orbits.}
    \label{fig:mloss}
\end{figure}

We estimate the mass lost by the system after each periastron passage using the usual energy criteria: a particle is considered unbound if its kinetic energy is higher than its gravitational energy,
\begin{equation}\label{eq:bound}
    E_\text{gr}+E_\text{kin} > 0.
\end{equation}
Altenatively, a fraction of the internal energy of the gas, $\alpha E_\text{int}$, can be converted into kinetic energy, and the amount of unbound particles is then determined as
\begin{equation}\label{eq:boundth}
    E_\text{gr}+E_\text{kin}+\alpha E_\text{int} > 0.
\end{equation}
Since $\alpha$ is unconstrained, setting it to unity allows us to derive an upper limit on the mass loss of the binary, while the more conservative criteria set by equation~(\ref{eq:bound}) can serve as a lower limit. Using both criteria, we evaluate the evolution of the mass loss throughout the simulations, which we show in Fig.~\ref{fig:mloss}.

During the first three orbits, the mass loss is very episodic, with a sharp rise in total unbound mass of $\sim$\,$3\times10^{-4}M_\odot$ right after each grazing interaction. Between interactions, the amount of unbound mass stays essentially constant. Most of the ejected particles have been cooled, radiatively accelerated, and unbound within a year from the periastron passage of the companion. Then, from the fourth orbit onward, the sudden rise in mass loss becomes stronger and the unbound mass is rising continuously. Both of these features intensify after each subsequent interaction. Since the surface layers of the RSG expand in response to the grazing interaction, and the companion reaches deeper and denser layers of the envelope, it drives more massive episodic outflows at periastron, but also a low-intensity continuous outflow until the next interaction. About $10^{-3}\,M_\odot$ of mass is unbound $1-2$~years after the interaction, and about $10^{-4}\,M_\odot$ is lost continuously in between each periastron passage. After the 10$^\text{th}$ orbit, the system approaches CEE and episodically ejects $\gtrsim10^{-2}\,M_\odot$ per interaction. The companion is completely engulfed by the envelope after the 13$^\text{th}$ orbit, and ejects about $0.1\,M_\odot$ of matter in a continuous way. At the end of the simulation, which corresponds to a total of 14 orbits, the system has ejected a total of $0.185\,M_\odot$, 80\% of which was ejected in the last ten years of the evolution, during the onset of CEE.

We note that our estimation of the mass ejected during the late evolution of our system is flawed due to our simplified treatments of cooling and dust condensation. Our treatment of radiative cooling does not take gas opacities into account since it relies on the approximation that the ejected gas is optically thin. However, this assumption does not hold for the outflows launched when the system is close to the onset of CEE, which should be optically thicker considering their high density. The opacity of these outflows likely decreases as they expand, but they will become optically thin and start to radiatively cool at much larger distances from the system than earlier outflows. Therefore, our method underestimates the temperature of the late outflows, allowing dust to condense and radiation pressure to drive winds in regions where it is not likely to happen. This leads to an overestimation of the outflow velocities and possibly of the amount of unbound mass at late times. 

Fig.~\ref{fig:mloss} also shows that the differences between the two mass loss criteria are relatively insignificant, especially at early times when the two criteria yield practically the same result. This is expected, since the internal energy of the gas at the stellar surface is much lower than the kinetic energy imparted to the gas by the companion during the interaction and any extra thermal energy from shock interaction is quickly radiated.
After five orbits, however, the companion starts to probe deeper layers of the envelope, with increasingly higher internal energy and where our cooling prescription is not active, so the two estimates deviate from one another. As the end of the simulation approaches, the gap between the two values deepens and we expect the difference between the two criteria to be similarly significant during the ensuing CEE phase.

\subsection{Evolution of the ejecta}
\label{sec:ejecta_evolution}

\begin{figure*}
    \centering
    \includegraphics[width=\linewidth]{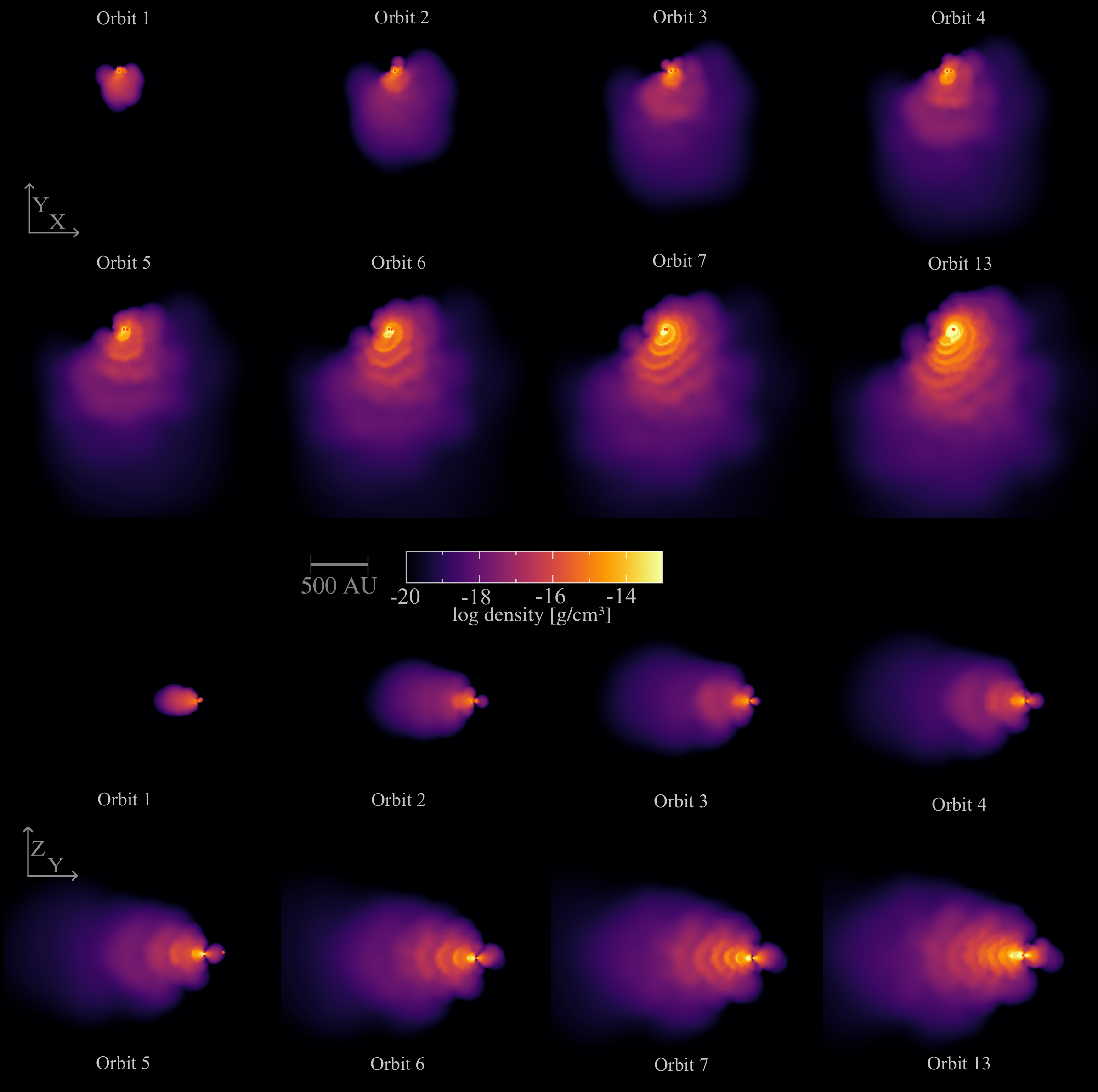}
    \caption{Density snapshots of the simulation with $2\times 10^6$ particles taken when the companion is at apastron. The two upper row show density cross section of the equatorial plane (slice along $z=0$) and the two bottom rows show the meridional plane (slices along $x=0$).}
    \label{fig:snaplong}
\end{figure*}

\begin{figure}
    \centering
    \includegraphics[width=\columnwidth]{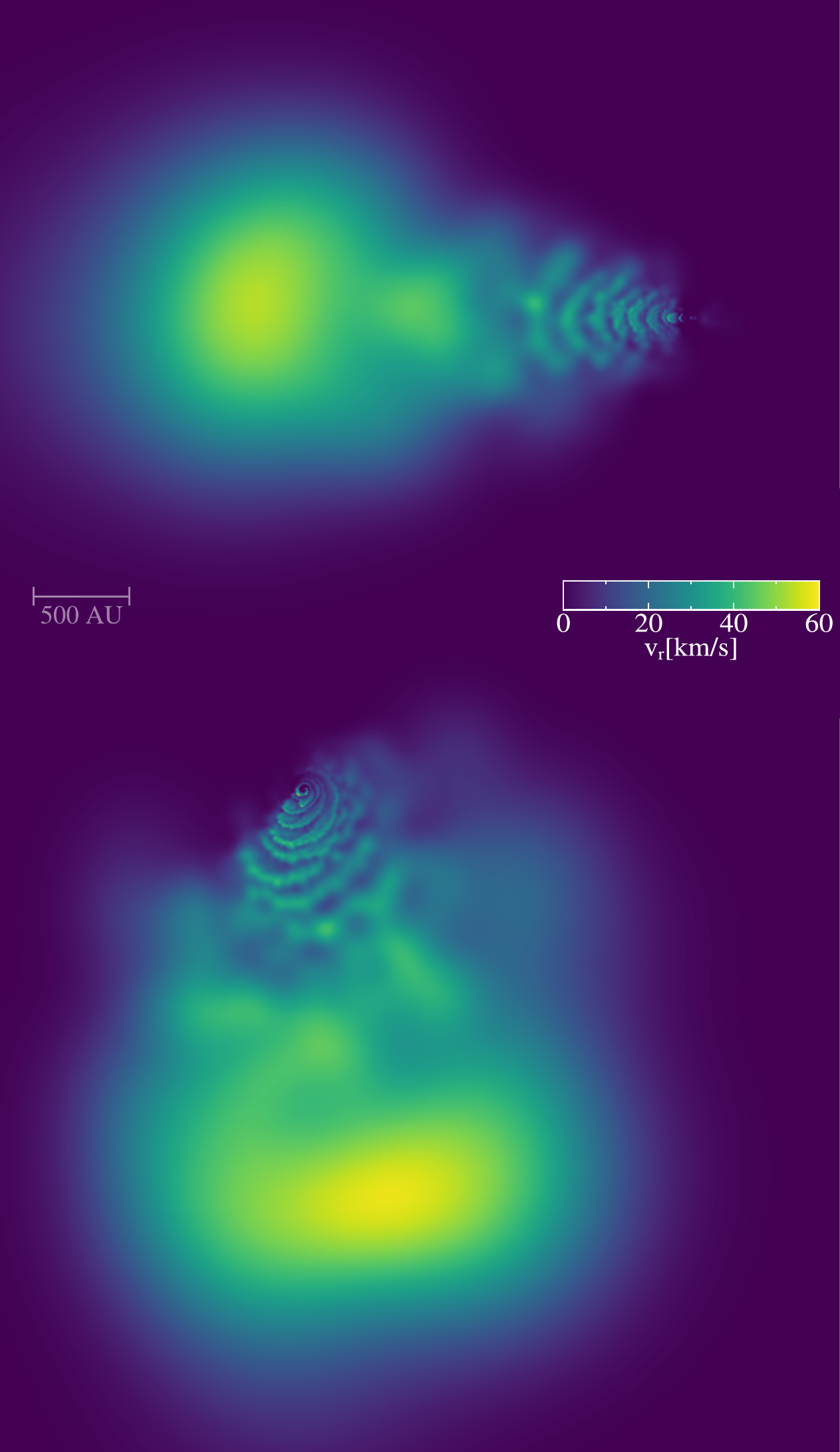}
    \caption{Snapshot of the radial velocity of the ejecta taken after 14 orbits. Upper and lower panels show radial velocities taken at $x=0$ and $z=0$ planes, respectively.}
    \label{fig:vel}
\end{figure}

In Fig.~\ref{fig:snaplong}, we track the evolution of the ejecta on longer timescales by showing density snapshots of our highest resolution simulation during subsequent orbits at the moment of apastron. Each successive periastron passage of the companion drives a new outflow, which first increases the density of the inner ejecta, then expands and merges with the less dense outer part of the ejecta. In this section, we use the term ``ejecta'' to qualify the entire body of ejected gas and the term ``outflow'' for the gas ejected due to one grazing interaction only. 

The evolution of the ejecta can be broadly divided in two phases that mostly differ by the time elapsed between subsequent grazing interactions. For the first five orbits, the orbital period ranges from 40 to 20 years and the outflows can expand over large scales before the binary is at periastron again, which causes the broad spiral pattern in the ejecta appearing in the first row of Fig.~\ref{fig:snaplong}. This slow evolution lasts for about 140~years, during which the ejecta expands to $r\gtrsim1000$\,au as a roughly semi-circular slab of vertical thickness $\sim$500\,au, with densities ranging from $10^{-15}$~g~cm$^{-3}$ in the innermost region to $10^{-19}$~g~cm$^{-3}$ in the outer ejecta. 

As seen in Fig.~\ref{fig:sep}, each interaction strengthens the orbital decay of the binary, which starts to strongly affect the outflows after the fifth orbit the binary. At this point, the grazing interactions become more frequent, with a period $P_\text{orb}\leq20$~yrs, causing tighter spiral patterns in the inner ejecta while the ejecta continues spreading outwards. The companion also starts to dig deeper into the envelope of the RSG, significantly increasing the density of each subsequent outflow, reaching up to $10^{-13}$~g~cm$^{-3}$ in the innermost part of the ejecta. As the system approaches the onset of CEE, we also see that matter is ejected more isotropically, and the densest part of the outflow spreads around the binary with a shape similar to a disc. We expect that the outflows will become even more isotropic as the system evolves through CEE, since the orbit should circularise, but the already ejected material should retain its overall asymmetry.

At the end of the simulation, the dusty ejecta broadly resembles a cone and we can distinguish three main regions. First, the innermost region, directly surrounding the binary, results from the late time evolution during which the binary was close to the onset of CEE and the outflows are the most massive. With a total mass of about $0.15\,M_\odot$ extending up to $\sim$100\,au away from the binary in a roughly conical shape, it is by far the densest region of the ejecta. Since the time elapsed between grazing interaction has drastically decreased, the shell-like over-densities caused by the companion shocking the gas are closer to each other, yielding a tighter spiral pattern than in earlier stages of the simulation. Beyond the innermost ejecta, we identify a relatively dense region of $10^{-15}-10^{-17}$~g~cm$^{-3}$ spreading up to $\sim$500\,au from the system, which we estimate to carry $\sim$$3\times10^{-2}\,M_\odot$. This region is more asymmetric than the inner ejecta, it corresponds to where most of the outflows have merged, and is thus made of the most extended parts of the late evolution outflows and the slowest regions of earlier outflows. Finally, a wide and less dense region extends to $r\gtrsim2000$\,au, with densities reaching very low values of $10^{-18}-10^{-20}$~g~cm$^{-3}$. This region carries roughly $5\times 10^{-3}\,M_\odot$, corresponding to the early outflows which have now spread far away from the binary.

In Fig.~\ref{fig:vel}, we show the radial velocity of the gas at the end of the simulation. The late outflows can be traced in the spiral pattern by sharp velocity fronts, while the early outflows have softer velocity gradients as they have merged long time ago. The outermost ejecta, arising from the first outflows and spreading beyond $r\gtrsim500$\,au, show radial velocities of $v_\text{r}\sim50-60$~km~s$^{-1}$. The rest of the ejecta has not reached such high velocities, the innermost part has on average $v_\text{r}\lesssim30$~km~s$^{-1}$ while the intermediate region reaches $v_\text{r}\approx40$~km~s$^{-1}$. We therefore estimate the wind terminal velocity to be $v_\infty\approx 60$~km~s$^{-1}$.

\section{Discussions} \label{sec:discussion}

Here, we discuss constraints on the evolutionary pathways that lead to the grazing by the companion (Sec.~\ref{sec:formation}), statistics of binarity of RSGs (Sec.~\ref{sec:rsg_binarity}), the duration of the grazing phase (Sec.~\ref{sec:duration}), what happens after the interaction (Sec.~\ref{sec:fate}), observational signatures (Sec.~\ref{sec:signatures}), and possible future improvements of our model (Sec.~\ref{sec:improvements}).

\subsection{Formation of the system}
\label{sec:formation}

The evolutionary pathways leading to our configuration depend on whether the RSG companion is a compact object, such as a neutron star, or a low-mass non-degenerate companion. From the point of view of our simulations both options are indistinguishable, because the gravitational potential of the companion must be smoothed on scales larger than its radius. If the companion is a low-mass main sequence (MS) star, the binary could have simply been born on this orbit and began interacting when the more massive star expanded to a RSG. Alternatively, the binary has such a low mass ratio that it may have formed in a wider more circular orbit and was subject to the Darwin and/or eccentric instability \citep[e.g.,][]{darwin_determination_1879,hut_stability_1980,eggleton01,pesta.2023}, which would have reduced the binary separation and potentially increased its eccentricity. Considering that VY~CMa is associated with but lies off of a cluster NGC~2362 \citep{lada78,melnik09,zhang_distance_2012}, it is interesting to speculate about a scenario where a massive star in a binary explodes as a supernova, leaves behind a neutron star, and the explosion kick sets the neutron star on an eccentric orbit and the binary on a trajectory away from its birth cluster. While this scenario might not be applicable to VY~CMa specifically, it is interesting to discuss it more generically. 

We consider a common scenario for the formation of double neutron star binaries \citep{tauris_formation_2017}, starting with two massive MS stars on a relatively close orbit. The more massive star eventually evolves and expands, and the binary experiences a phase of case B or C mass transfer.
As the RSG transfers mass to its companion, it is gradually stripped of its envelope and eventually undergoes core collapse (CC). The collapsing star turns into a NS without fully disrupting the binary, which according to \citet{renzo_massive_2019} should happen in about 14\% of massive binaries. While the natal kick of the NS does not fully unbind the binary, it will increase its eccentricity \citep{brandt_effects_1995, kalogera_orbital_1996} and widen the orbit to up to 4 times the pre-CC separation according to \citet{kalogera_orbital_1996}. The initially less massive star will expand to a RSG, eventually reaching radii large enough for the compact companion to graze the RSG envelope at periastron. 

The timescale of the post-CC evolution depends on how close the initial masses of the two stars were and whether the accretion during the pre-CC mass transfer phase has increased the main sequence lifetime of the accretor \citep{neo_effect_1977}. It is, however, strictly constrained by the tidal circularisation timescale of the binary \citep{zahn_tidal_1977}, since the star should reach the RSG phase before the eccentricity of the binary decreases significantly. If we consider that the initially less massive star is close to the onset of the RSG phase at the birth of the NS, and therefore has a significantly convective envelope, the circularisation timescale of the binary is highly dependent on the ratio of semi-major axis to stellar radius $\tau_\text{circ}\propto (A/R_1)^8$. Using Eq.~2 of \cite{verbunt_tidal_1995}, the circularisation timescale reaches a maximum estimate of $\tau_\text{circ}\simeq10^{7}$yrs for a $20\,M_\odot$, $1500\,R_\odot$ RSG. The binary should thus circularise on longer timescales than the timescale of the evolution and expansion of the future RSG, which should at most of the order of $10^6$~yrs if the star is still on the MS at the start of the post-CC evolution. Therefore, the binary should retain its large eccentricity by the time the compact companion can graze the envelope of the RSG.

\subsection{Statistics of RSG binarity}
\label{sec:rsg_binarity}

Statistics of RSG binarity can serve as a clue whether our scenario occurs frequently.
\cite{neugent_runaway_2018} devised some criteria to detect companions around RSGs from their contribution in the spectra of the binaries, however, they are mostly efficient for blue companions since their contamination is more easily distinguished from the red light of the RSG. A complementary way of determining the binarity of a RSG is to secure multi-epoch spectroscopy to look for periodic variations of radial velocities. However, due to the large radii of RSGs the minimum orbital period of their companion has to be of the order of hundreds of days, meaning that such spectroscopic studies have to be performed on timescales of years. Furthermore, the amplitude and timescale of these periodic variations are likely similar to the variability due to atmospheric convective motion in RSGs \citep{schwarzschild_scale_1975}, which further complicates RSG multiplicity studies. Despite these difficulties, the binary fraction of cool supergiants in the Milky Way has been estimated to around 35\% \citep{burki_nineteen_1983}, and some X-ray binaries were found to have a RSG companion \citep[e.g.,][]{gottlieb_rapidly_2020, hinkle_m_2020}. RSG multiplicity studies targeting the Local Group (e.g., \citealt{neugent_binary_2019,patrick_vlt-flames_2019,neugent_red_2020,dorda_multiplicity_2021,neugent_red_2021}) generally find binary fractions around 30\%, mostly with OB-type companions. \cite{patrick_red_2022} found 6 candidates for RSG~+~compact companion in the Small Milky Cloud, though they potentially are  false positives, and \cite{neugent_red_2021} estimated the fraction of RSG~+~compact companion in M31 and M33 to be about 4.73\% using BPASS \citep{eldridge_binary_2017}. While rates of binarity for RSGs inferred from observations are relatively low, especially for binaries with a compact companion, it is still non negligible and therefore our grazing scenario seems plausible. 

\subsection{Duration of the grazing encounters}
\label{sec:duration}

Our simulations cover only about 200 years or 13 orbits before the binary enters the CEE. Such a short duration would imply that our chances of finding a RSG undergoing this type of evolution is very slim given the RSG lifetime. However, the duration of our simulation is mostly driven by the constraints of available computing time and resolution. If both can be increased, we could simulate the evolution for many more orbits before the CEE. This interaction is similar to a normal CEE, where \citet{iaconi18} and \citet{reichardt19} found that increasing simulation resolution leads to longer duration of pre-CEE inspiral, mass transfer, and mass loss. Long duration of pre-dynamical CEE phase is also seen in 1D binary evolution simulations \citep[e.g.,][]{klencki21,marchant21}. Observations of transients accompanying stellar mergers, the luminous red novae, also indicate the presence of pre-dynamical mass-loss lasting many hundreds and potentially thousands of binary orbits \citep[e.g.,][]{tylenda11,pejcha16a,pejcha16b,pejcha17,blagorodnova_luminous_2021}.

Another aspect affecting the duration of the grazing phase is missing physics near the surface layers of the star. Real stars exhibit a complex interplay of convection, diffusion, and ionization in their surface layers, which affects the stability with respect to mass removal \citep[e.g.,][]{pavlovskii15}. In RSGs specifically, the additional physical effects include large size of convective cells, pulsations leading to in the outer atmosphere shocks, dust and molecule formation in the gas lifted off the surface by pulsations, and feedback from accretion onto the companion \citep[e.g.,][]{haubois09,shiber18,goldberg_numerical_2022,freytag_global_2023}. 

All of these effects could extend the duration of the grazing phase by making the conditions in the outer stellar layers different from what we assume in our adiabatic simulations. Furthermore, the mass ejections could happen intermittently depending on whether the periastron passage occurs during the maximum or minimum expansion of the RSG pulsation. Finally, longer duration of the grazing phase could also facilitate precession and tilting of the companion orbit, for example, due to action of tidal forces. The mass ejections would then still be oriented in one direction during each ejection or a series of subsequent ejections, but the orientation of ejection could gradually change. All of these aspects would be interesting to investigate in future work.

\subsection{Evolution and fate of the binary}\label{sec:evol}
\label{sec:fate}
Depending on the amount of envelope ejected during the CEE ensuing after the grazing and the nature of the companion, the system could evolve to a close binary with two compact objects (e.g., a double NS binary, \citealt{tauris_formation_2017}), or the two stars could merge, possibly resulting in a Thorne-\.{Z}ytkow object (ZTO) \citep{thorne_stars_1977} if the companion is a NS, or an exotic supernova if the merger product explodes \citep[e.g.,][]{chevalier12}. It is, however, not possible to conclude on the outcome of this process from our simulations, since properly simulating this phase of the evolution of the binary would require to resolve the deep interior of the RSG. 

We may broadly examine the outcome of the CEE using the energy formalism \citep[e.g.,][]{webbink.1984}:
\begin{equation}
-G \frac{M_1 M_{1, \text{e}}}{\lambda R_{1}}=-\alpha_\text{C E} G\left[\frac{M_{1, \text{c}} M_2}{2 A_\text{f}}-\frac{M_1 M_2}{2 A_\text{i}}\right]
\end{equation}
where $M_{1, \text{e}}$ and $M_{1, \text{c}}$ are the mass of the RSG envelope and core respectively, which are about $13.5\,M_\odot$ and $6.5\,M_\odot$ according to the \textsc{MESA} model used to setup our RSG. Parameter $\lambda$ is defined by the mass distribution inside the star, $A_i$ and $A_f$ and the initial and final separation of the CEE. By setting $\alpha_\text{CE}$, the envelope ejection efficiency, and $\lambda \lesssim 0.1$ as appropriate for evolved red supergiants \citep{kruckow16}, we find that the CEE can unbind the whole envelope for final separation of $A_f\lesssim4\,R_\odot$.  Thus, it seems unlikely that the system ejects the whole envelope. This estimation is however too simple to reflect the reality, since the efficiency of envelope ejection in CEE and the physical processes involved (e.g., recombination, dust formation) are still relatively unconstrained in current CEE models (see \citealt{ropke_simulations_2023} for a detailed discussion). Therefore, we leave the fate of the binary as an open question.

We can, however, estimate the impact of our chosen initial conditions on the evolution of the system. We could consider a companion mass $M_2$ lower than $2\,M_\odot$, in which case we expect less massive outflows due to the grazing interaction. In our simulations, the efficiency of the radiative force to drive the winds would remain unchanged due to the very simple formulation of radiative pressure on dust grains, therefore we do not expect the wind velocities to change for a different companion mass. However, we note that in a realistic scenario the efficiency of dust condensation does depend on the density of the gas, and since a lower mass companion yields less dense outflow, it would also likely change the wind properties. The decrease in companion mass would also cause the orbit to decay more slowly, as the dynamical friction between the outer envelope and the companion is lower during the each periastron passage, which will result in a later plunge-in. We expect opposite effects for higher companion masses: more massive outflows and a faster evolution.

The choice orbital parameters of the system should also have a significant influence on the evolution of the binary and the outflows. In our simulations, $a$ is not freely chosen, it is set so that the binary separation at periastron is exactly equal to the radius of the RSG, $a=R_1/(1-e)$. For lower $e$, $a$ is smaller and the less eccentric orbit will be overall closer to the RSG surface. As a result, the companion graze a larger portion f the stellar surface and encounter higher densities. This leads to less asymmetric outflows, and the higher drag on the companion accelerates the orbital decay to CEE.

\subsection{Observational signatures}
\label{sec:signatures}
Here, review various observations that can indicate that grazing encounters are ongoing or happened relatively recently: outflow emission, changes of the RSG, accretion onto companion, observable signatures of TZO, and supernova explosions. We also specifically discuss VY~CMa.

\subsubsection{Outflow emission}
The most evident signature of this interaction is the emission from the cool asymmetric outflow, which we expect to be observable at millimetre/sub-millimetre and infrared wavelengths. As shown in Fig.~\ref{fig:snaplong} the dusty outflow could be angularly resolved with the ALMA interferometer if the distance to the system is small enough. Outside of the continuum emissions from the dust, the ejecta should also be traceable with molecular emissions, which arise from the various gas-phase chemistry across the outflow. Since we do not model the various phenomena occurring at the RSG surface, the morphology of the outflow in our simulations represents the simplest possible outcome. However, any complicated morphological feature in the outflow would likely be unresolvable, which includes the spiral features in the innermost ejecta, and traces of the common envelope phase would likely be hard to detect.

\subsubsection{Changes of the RSG}

Repeated close passages of a companion will affect the RSG by tidal dissipation. Our simulations are inadequate to study these processes, but we can speculate about some of the potential outcomes. First, the RSG could expand in response to the tidal heating. For our choice of companion mass, the orbital energy of a grazing orbit is much smaller than the RSG envelope binding energy \citep{klencki21} and no significant overall expansion is expected. However, the dissipated energy could still strengthen the processes that are already ongoing such as the more common dust-driven RSG wind, potentially leading to a tidally-enhanced wind \citep[e.g.,][]{tout88,chen11}. The intensity of this effect depends on the amount of tidal dissipation, the depth of deposition, and the ability of the star to quickly remove this excess energy by radiation.

\subsubsection{Accretion onto companion}

Before fully entering the envelope of the RSG, our system should also show X-ray emission from the accretion on the companion. Since the companion moves supersonically through the outflow of the RSG, we can approximate the accretion rate $\dot{M}_2$ of the companion using the approximation of spherical accretion onto a moving object (Bondi-Hoyle-Lyttleton accretion),
\begin{equation}\label{eq:accrate}
\dot{M}_2 = \frac{4\pi \rho (GM_2)^2}{(c^2_\text{s} + v_2^2)^{3/2}}   ,
\end{equation}
where $\rho$ is the density of the surrounding outflow, $c_\text{s}$ is the sound speed, $M_2$ the mass of the companion, and $v_2$ is the velocity of the companion relative to the surrounding gas. In our simulations, the average relative velocity of the companion is $~20$~km~s$^{-1}$, and the density of the surrounding ejecta is $~10^{-15}$~g~cm$^{-3}$, resulting in an accretion rate of $4 \times 10^{-6}\,M_\odot$. The corresponding accretion luminosity is $L_\text{acc} = GM_2\dot{M}_2/R_2$ where $R_2$ is the radius of the companion. Inefficient accretion or radiative emission will decrease $L_\text{acc}$. When considering a low mass MS companion of $R_2\simeq2R_\odot$, the accretion luminosity becomes $L_\text{acc}\simeq 10^{35}$~erg~s$^{-1}$, making it similar to X-ray bright T Tauri stars \citep{telleschi07}. For a NS companion of $R_2=10$~km~s$^{-1}$, the accretion rate yields a very high luminosity of $L_\text{acc}\simeq 10^{40}$~erg~s$^{-1}$, which is a factor of $\sim$50 higher that the Eddington limit for a $2\,M_\odot$ star. Given the super-Eddington accretion rate, the NS might look like an ultra-luminous X-ray source displaying super-Eddington flux only near the polar axis \citep[e.g.,][]{king09}. If jets develop, our scenario reduces to the grazing envelope evolution developed by \citet{soker_close_2015},  \citet{shiber17}, and \citet{shiber18}. 

The high gas density in the vicinity of the RSG that gives rise to a high $\dot{M}_2$ also provides a large gas column that can absorb and scatter the X-rays. Considering a spherical constant-velocity outflow from the RSG with $r^{-2}$ density profile, $\dot{M}_1 = 10^{-5}\,M_\odot$\,yr$^{-1}$, velocity 20\,km\,s$^{-1}$, and inner edge at $1500\,R_\odot$, we obtain density $\approx 10^{-15}$~g~cm$^{-3}$ and hydrogen column of $\sim 10^{23}$. Depending on the emission temperature and the detector properties, such a high gas column can reduce the expected flux by many orders of magnitude \citep[e.g.,][]{montez_constraints_2015}. Even with inefficient accretion and high intervening gas column, we would still expect signatures of accretion such as highly ionized gas or feedback on the surrounding medium.

\subsubsection{Observable signatures of TZO}

As we discussed in Sec.~\ref{sec:evol}, the companion could have spiraled in and merged with the core of the RSG, possibly resulting in a TZO. These objects are hard to distinguish from RSGs, only a few candidates have been proposed, such as U~Aqr \citep{vanture_is_1999}, HV~2112 \citep{levesque_discovery_2014,tout_hv2112_2014,maccarone_large_2016,worley_proper_2016,beasor_critical_2018}, and VX~Sgr \citep{tabernero_nature_2021}, and none of these objects have been confirmed to be TZOs. According to \cite{cannon_structure_1992}, a TZO should look like a cool and bright RSG close to the Hayashi limit \citep{hayashi_outer_1961}, but show unusual lines in its optical spectra. As the neutron star spirals in the envelope of its companion, the combination of the high surface temperature of the NS and convective envelope of the RSG would trigger interrupted rapid proton (irp) capture \citep{cannon_massive_1993}. Despite the broad TiO absorption features observed in RSGs, some of the heavier elements produced by the irp processes should have high enough abundances in the atmosphere to be observed, especially Rb, Yb, and Mo \citep{biehle_observational_1994}. Other TZO spectral features include  $^7$Li and Ca~I enhancement \citep{podsiadlowski_evolution_1995, tout_hv2112_2014} or $^{44}$TiO and $^{44}$TiO$_2$ \citep{farmer_observational_2023}.

\subsubsection{Supernova explosions}

The merger of the RSG core and the companion may also result in a supernova explosion, which could produce a black hole remnant if the companion is a compact object. Many Type IIn supernovae and other luminous transients show evidence for mass ejections preceding the terminal supernova explosion \citep[e.g.,][]{ofek14,margutti14,jacobsongalan22}. It has been suggested that a strong binary interaction or a CEE is responsible for pre-supernova mass ejections leading to a formation of dense circumstellar medium \citep[CSM, e.g.,][]{smith11,chevalier12,metzger22}. This has motivated theoretical works investigating interactions of supernova explosions with various aspherical CSM distributions such as disks, oblate or prolate ellipsoids, and colliding winds shells \citep[e.g.,][]{vlasis16,suzuki19,kurfurst20,pejcha22}. Interestingly, none of these works have considered the type of aspherical CSM that we are predicting here: subtending only a small fraction of a solid angle with internal density variations corresponding to individual periastron passages. However, based on analogous works on different aspherical CSM distribution, we can predict that supernova explosion colliding with the CSM predicted here would lead to a radiative shock that is quite likely embedded in the optically-thick supernova ejecta and that initially reveals its presence only as an additional energy source. When the supernova ejecta becomes optically-thin, the CSM distribution could manifest in profiles of nebular spectral lines. 

\subsubsection{Application to VY~CMa}

We can compare these expected signatures to observations of VY~CMa. We first note that no companion has been directly observed around VY~CMa, suggesting that if our scenario was ever involved in the evolution of VY CMa, the companion is now either fully engulfed in the envelope of the giant, or is orbiting too close to the stellar surface to be observable
\citet{decin06} reconstructed the mass-loss history of VY~CMa and found that it underwent a phase of increased mass loss rate of $\simeq 3\times 10^{-4}\,M_\odot$\,.yr$^{-1}$ about 1000 years ago. This phase lasted about 100 years and was preceded by a phase with relatively low mass loss rate of $\simeq 10^{-6}\,M_\odot$\,yr$^{-1}$ for about 800 years and succeeded by a phase of increased mass loss of $\simeq 10^{-4}\,M_\odot$\,yr$^{-1}$ lasting until today. It would be natural to identify this event of increased mass loss with grazing interactions, which culminated with a stellar merger followed by an enhanced wind phase due energy deposition in the RSG envelope by the inspiralling companion \citep[e.g.,][]{clayton17,glanz18}. 

Alternatively, the morphology of our asymmetric outflow seems to be consistent with the dusty clumps observed in the immediate vicinity of VY~CMa \citep[e.g.,][]{kaminski_massive_2019}. The mass of these clumps was estimated to be of the order of $10^{-3}-10^{-2}\,M_\odot$ with velocities of $~20-50$~km s$^{-1}$, and were likely ejected about 100 years ago \citep{humphreys_hidden_2024}, which is consistent with the total mass of the extended ejecta in our simulations and the velocities shown in Fig.~\ref{fig:vel}. We note that our wind terminal velocities only depend on the Eddington factor, which is a free parameter set in equation~(\ref{eq:gamma}), so they are a broad upper limit rather than a reliable estimate. \citet{kaminski_massive_2019} estimated the size of the clumps with 3D radiative transfer models, and found that the most elongated clump (clump B) could be up to $1000$\,au long, which is compatible with the size of our ejecta.

Concerning the companion accretion, \cite{montez_constraints_2015} obtained non detections in X-rays that place an upper limit of $L_\text{X}<1.6 \times 10^{31}$~erg~s$^{-1}$, which is far below the estimated $L_\text{acc}~10^{35}$~erg~s$^{-1}$ we derived. However, the upper limit is highly contingent on the emission temperature and the sensitivity to softer radiation with $T\lesssim 10^6$\,K is much worse given the expected intervening absorption column.  Still, VY~CMa does not prominently display highly ionized emission lines or other signs of an accretion process. So if VY CMa had a companion, it was probably already deep within the RSG envelope by the time the observations were made. If the companion is a NS then we might expect VY~CMa to be a TZO, spectroscopic studies of the object do show Ca I and Rb I lines \citep{wallerstein_spectroscopic_1971, dinh-v-trung_high-resolution_2022}, however, no traces of other heavy elements have been found. The use of different tracers, such as the ones proposed by \cite{farmer_observational_2023}, could shed some light on this issue. 

To summarize, VY~CMa could have merged with a grazing companion about 1000 years ago, however, the complexity of its surrounding medium would require some modification to our model or combination with other physical processes. Increasing the realism of our model could  allow to better determine whether this mechanism plays a role in the formation of some of the asymmetric outflows around VY~CMa.

\subsection{Improvements of the model}
\label{sec:improvements}

Our simulations use approximations and simple treatments of the involved physical processes to reduce their computational cost. Here, we discuss potential improvements for follow-up studies.

Firstly, we have made several assumptions regarding the initial conditions of our setup. We have assumed that when the simulations start, no interactions are taking place between the RSG and the companion. Considering how close the two stars are, there is a chance that the RSG overfills its Roche Lobe around the periastron of one or more orbit preceding the start of the simulations, resulting in potential mass transfer and/or outflows. However, any interaction during the previous orbits would simply lead to smaller, possibly negligible, mass ejections. To assess the relevance of previous interactions, one could evolve the binary in 1D in MESA following the methods of \cite{marchant21}, and possibly obtain better initial conditions for our simulations. 

Another approximation we made in this study is to model the RSG envelope as convectively stable, which prevent us from investigating the interplay between surface convection and the companion interaction, which would likely affect the morphology of the ejecta. However, resolving convective motion requires to model the entire RSG envelope as well as including radiative transport in the envelope, to avoid a strong overestimation of the convective flux (e.g., \citealt{ricker_common_2018}). Both choices would dramatically increase the cost of the simulations, especially considering the long timescale of our simulations, therefore, incorporating convection in our simulations will likely be considered in later phases of improvements.

Furthermore, as mentioned in Sec.~\ref{sec:cooling} and \ref{sec:dust}, our numerical treatment of dust-driven winds has been simplified and can be improved to various degrees. Dust formation and destruction can be more accurately treated by including a density and composition dependence, for instance by using the moment method \citep{gail_dust_1984,gail_dust_1988,gauger_dust_1990,gail_physics_2013}, which has been implemented for carbonaceous dust grains in \textsc{Phantom} by \cite{siess_3d_2022}. RSG stars have a low C/O ratio, so this method would need to be adapted to oxygen-rich dust condensation, which is significantly harder to model than carbon-rich dust formation. There are much more dust species to account for compared to the case of carbon-rich dust, and it is not completely known which particles serve as seed nuclei for oxygen dust growth (for more details, see Chapter~15.5.6 of \citealt{gail_physics_2013}). Implementing a complete condensation scheme for oxygen-rich dust in 3D hydrodynamics codes is therefore a challenging but important step towards improving dust treatment in simulations of evolved stars. It would also allow for a better estimation of local opacities and Eddington factor, which are essential for the proper treatment of radiation pressure on the dust-gas mixture. 

We could further improve the realism by resolving the dust-gas interactions. \textsc{Phantom} already has several formalisms for dust-gas mixtures, either with a two-fluid approach \citep{laibe_dusty_2012} or a one fluid method that keeps track of the composition of the mixture \citep{laibe_dusty_2014,laibe_growing_2014,price_fast_2015}. While the outcome of simulations with accurate treatments of dust-driven winds might be qualitatively similar to our results, it would also likely yield a more detailed structure in the winds, and lower outflow velocity and densities since our dust condensation criteria is relatively permissive.

Improvements regarding the formulation of the radiative force on dust grains would also likely impact the morphology of the wind. \cite{esseldeurs_3d_2023} shows the difference in radiative pressure treatments for winds in binaries with an AGB star, and it is clear that more accurate approximations (e.g, \citealt{lucy_formation_1971,lucy_mass_1976}) yield more detailed wind structures than with a simple free-wind approximation, as well as better estimations of the wind velocities. However, such methods  require the calculation of optical depth, which needs to be estimated along the line of sight of each particles and severely complicates the simulations if one wants to calculate it on-the-fly during an SPH simulations. This was implemented in \textsc{Phantom} by \cite{esseldeurs_3d_2023} using a ray tracer algorithm, so we leave the possibility of adapting their method to our setup for follow-up studies.

\section{Conclusions}
The goal of this work was to investigate whether a companion grazing the envelope of a RSG can launch significant asymmetric episodic outflows that later expands through dust-driven winds. To do so, we performed 3D hydrodynamics simulations of a $2\,M_\odot$ star on a highly elliptical orbit around a $20\,M_\odot$ RSG with an envelope extending to $1500\,R_\odot$ (Fig.~\ref{fig:setup}, Sec.~\ref{sec:methods}). In our models, we see the companion grazing the RSG envelope at periastron and ejecting gas from the outermost envelope, which results in a dense semi-circular outflow (Fig.~\ref{fig:snapclose}). The ejected gas becomes optically thin and cools, reaching conditions that are favourable for dust condensation. The radiative pressure on dust grains then accelerates the outflow outwards, effectively launching dust-driven winds that expand radially.

We investigated the evolution of the system through several successive grazing interactions, and found that the orbit drastically tightens after each interaction (Fig.~\ref{fig:sep}). The orbital period decreases by $\sim3-4$~years per orbit, decreasing the periastron distance significantly and enhancing the mass loss during the grazing interaction. The outflows therefore become denser and more frequent as the system evolves, effectively altering the properties of the ejecta (Fig.~\ref{fig:snaplong}). The mass ejected during each grazing interaction goes from $3\times10^{-4}\,M_\odot$ during to first orbit to  $\sim10^{-2}\,M_\odot$ before the onset of CEE (Fig.~\ref{fig:mloss}).

After 13 orbits ($\sim$200~years), the system enters CEE which dramatically enhances the orbital evolution and mass loss rate of the binary. Due to the large softening length of the numerical core of the RSG, we cannot resolve the CEE of the system with our simulations, therefore the outcome of the CEE remains unconstrained and could result in a short-period binary or a merger. Our simulations stop after 14 orbits, at which point the binary has ejected a total of $0.185\,M_\odot$ of gas spreading beyond $r\gtrsim1000$\,au, with 80\% of this mass situated in the innermost part of the ejecta ($r\lesssim100$\,au). The final ejecta has a conical shape and shows a shell-like structure due to the shocks from each grazing by the companion.

While the initial conditions of our simulation seem relatively exotic, mostly due to the eccentricity of the system, we expect this grazing interaction to be relevant for the evolution of massive binary systems. For instance, this scenario is applicable to binaries with low mass MS companions with an orbit wide enough to retain a significant eccentricity, in which the stars evolve as effectively single until the massive star expands as a RSG or the orbit dramatically tightens due to the Darwin instability. 

Although the duration of this phase is uncertain and depends on complicated physics near the RSG surface, it should still produce observable signatures. The dusty ejecta should be observable at millimetre/sub-millimetre wavelengths, as well as molecular lines due to the rich chemistry expected in the outflow. 

Additionally, the companion should accrete matter from the outflows, which we expect to result in X-ray emission or highly ionized lines. However, their detectability is highly affected by the large intervening absorption column expected in such situations.

Lastly, the binary could evolve to a TZO after CEE if the companion is a NS, which is of interest since TZOs are hard to distinguish from RSG, and none have been unambiguously identified yet.

By comparing our results with observations of the ejecta around VY~CMa, we speculate that such a grazing interaction could have been responsible for the increase in mass loss occurring about 1000 years ago \citep{decin06} or more recent ejections about 100 years ago \citep{humphreys_hidden_2024}. Some outflows around VY~CMa resemble results of our simulations, however, the observed morphology is much more complex and requires an interplay of multiple effects.
Our simulations, which were only meant as a proof of concept, were performed using simple treatments of radiative cooling and winds in the outflows, and therefore require more accurate prescriptions in follow-up studies. Improving the accuracy of our simulations will allow us to study the morphology of the outflows in greater details, as well as produce synthetic observables that can used for a better comparison with observations of VY~CMa, providing a strong test for our hypothetical scenario.

\section*{Acknowledgements}
We thank the anonymous referee whose comments helped to improve this work. CL thanks Shazrene Mohamed for the fruitful discussions and suggestions and Mike Lau for the help on generating stellar profiles. This research has been supported by Horizon 2020 ERC Starting Grant ‘Cat-In-hAT’ (grant agreement no. 803158). The work of CL has been supported by the Charles University Grant Agency project No 116324. The work of OP has been supported by the Charles University Research program No. UNCE24/SCI/016. This work was supported by the Ministry of Education, Youth and Sports of the Czech Republic through the e-INFRA CZ (ID:90254). The simulations were performed using the Barbora cluster at IT4Innovations and the allocations provided by the projects OPEN-27-60 and OPEN-30-50.

Software: NumPy \citep{van_der_walt_numpy_2011}; Matplotlib \citep{hunter_matplotlib_2007}; Astropy \citep{astropy_collaboration_astropy_2013}; MESA \citep{paxton_modules_2011}; Phantom \citep{price_phantom_2018}, Splash\citep{price.2007}.

\section*{Data Availability}

The output files from our simulations will be shared on reasonable request to the corresponding author. The video of the snapshots of the first interaction is available at \href{https://youtu.be/jcW0KyMayBE}{https://youtu.be/jcW0KyMayBE}.



\bibliographystyle{mnras}
\bibliography{main} 




\appendix

\section{Red supergiant profile}\label{ap:rsgprof}
We need a realistic RSG profile to establish the boundary conditions (surface pressure and radius) and core size of our simple RSG interior. To obtain this stellar model, we use MESA v22.11.1 with the provided \textsc{20M\_pre\_ms\_to\_core\_collapse} test case to evolve a $20\,M_\odot$ zero-age main sequence star with metallicity Z=0.02 to the RSG phase without stellar winds. To produce a stellar model similar to VY~CMa, we need to ensure that the stellar radius is large enough to reach very low densities in the outer envelope. As such, we looked for stellar evolution parameters for which the star expands as much as possible during the RSG phase. Similarly to \cite{goldberg_numerical_2022}, we found that the largest envelopes result from low mixing length coefficient in the H-rich envelope $\alpha_H$, i.e. less efficient convection. More specifically, we found that a model with $\alpha_H=1$ reaches about 1150\,$R_\odot$ in the RSG phase while models with $\alpha_H=3$ only expand to 1000\,$R_\odot$.

We therefore choose to work with the model with the lowest mixing length and opt to analytically expand the  stellar profile obtained to reach a radius of 1500~$R_\odot$. This expansion is done following homology scaling relations, which are obtained by estimating the equations of stellar structure by including the conservation of mass distribution:
\begin{equation} \label{eq:massdistr}
    \frac{r_\text{i}(m)}{R_\text{i}}=\frac{r_\text{f}(m)}{R_\text{f}}.
\end{equation}
The new surface pressure of the star $P_f$ scales as follows:
\begin{equation} \label{eq:homologiesP}
    P_\text{f} = P_\text{i} \left ( \frac{M_\text{f}}{M_\text{i}} \right )^2 \left ( \frac{R_\text{i}}{R_\text{f}} \right )^4.
\end{equation}
We take the effective radius of the excised core to be half of the total radius of the star, which corresponds to a softening length of $r_\text{soft,i}=287.5\,R_\odot$ and a core mass of $M_\text{c}=13.75\,M_\odot$ according the the MESA stellar profile. The radius of the core of the extended giant $r_\text{c,f}$ is simply calculated using equation~(\ref{eq:massdistr}). 

\begin{figure}
    \centering
    \includegraphics[width=\columnwidth]{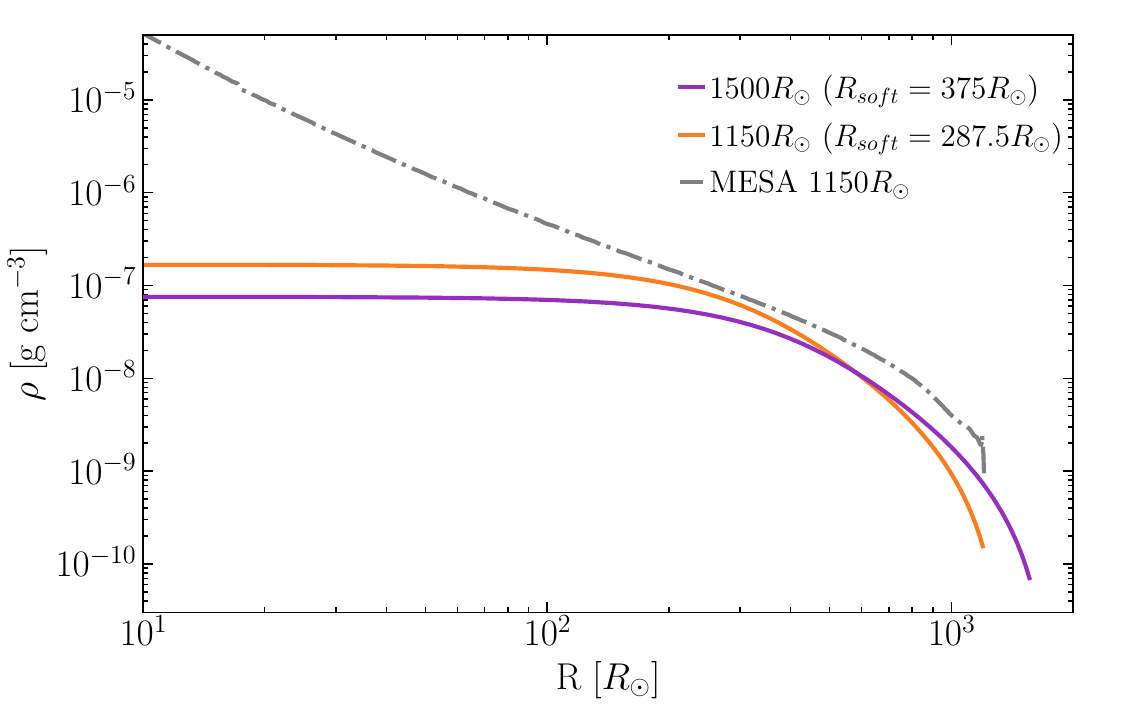}
    \caption{Comparison of the density profile of the RSG model obtained with MESA (grey line) with our simplified stellar profiles. The orange line shows the simplified profile calculated with the initial conditions from the MESA model, the purple line shows the profile calculated with the expanded initial conditions.}
    \label{fig:rsgext}
\end{figure}

Following this procedure, we expand a 1150\,$R_\odot$ star with $P_\text{i}=212$ dyn/cm$^2$ and $r_\text{soft,i}=287.5\,R_\odot$ to a 1500\,$R_\odot$ star with $P_\text{f}=73$ dyn/cm$^2$ and $r_\text{c,f}=375\,R_\odot$. These boundary conditions are then used to solve the equations of hydrostatic equilibrium to obtain a convectively stable stellar interior model as explained in Appendix~A of \cite{lau_common_2022}. To construct this model we use an ideal gas equation of state with adiabatic index $\gamma=5/3$ and a uniform mean molecular weight $\mu=0.659$ (obtained from the average mean molecular weight in the envelope of the MESA model). In Fig.~\ref{fig:rsgext}, we compare the detailed MESA model to the simple profiles obtained using initial conditions from MESA RSG and the extended extended initial conditions. Taken at face value, the density in the upper layers of the envelope is underestimated, and density gradient is severely smoothed. This would lead to an underestimation of the outflow density and the impact of the grazing on the companion trajectory. However, the actual density profile near and above the photosphere is affected by pulsations and wind launching and likely significantly differs from a simple 1D stellar model.

\bsp	
\label{lastpage}
\end{document}